\documentclass[preprint,aps]{revtex4-1}

\usepackage{graphicx}% Include figure files
\usepackage{dcolumn}% Align table columns on decimal point
\usepackage{bm}% bold math
\usepackage{latexsym}
\usepackage{amsfonts}
\usepackage{amssymb}
\usepackage{amsmath}
\usepackage[usenames]{color}
\begin{document}
\preprint{KUNS-2556}
\preprint{KOBE-TH-15-04}

\title{Primordial Gravitational Waves in Bimetric Gravity}
\author{Yuki Sakakihara}
\affiliation{Department of Physics, Kyoto University, Kyoto 606-8502, Japan}
\author{Jiro Soda}
\affiliation{Department of Physics, Kobe University, Kobe 657-8501, Japan}

\date{\today}% It is always \today, today,

\begin{abstract}
We study primordial tensor power-spectra generated during inflation in bimetric gravity.
More precisely, we examine a homogeneous expanding spacetime in a minimal bimetric model with
 an inflaton and calculate tensor perturbations on the homogeneous
 background under slow-roll approximation.
In terms of the mass eigenstates, only the power-spectrum of the massless state remains constant 
and both the power-spectrum of the massive state and the cross power-spectrum
 rapidly decay during inflation.
The amplitude of the physical power-spectrum is suppressed due to the flavor
 mixing.
All power-spectra in the flavor eigenstates coincide with each other 
up to the first order of the slow-roll parameter.
\end{abstract}


\maketitle

\tableofcontents

\section{Introduction}
Although no one doubts the existence of gravitons, no one has really observed them.
Indeed, we do not know if gravitons have mass or how many species there are.
What we can confidently say is at least one of them must be sufficiently light in order to realize
the Newtonian potential.
Thus, we have room to suppose two or more graviton species exist.
It is well known that two interacting massless graviton can not exist.
However, it was recently found that a massless graviton and a massive graviton
can exist at the same time \cite{deRham:2010ik,deRham:2010kj,Hassan:2011hr,Hassan:2011zd,Hassan:2011ea}.
The form of interaction terms is highly constrained in order to avoid ghosts.
Such theory necessarily includes another metric in addition to the
physical metric and is called bimetric gravity.
In general, the interaction terms include five theoretical parameters.

We are interested in if the bimetric theory is theoretically consistent and reconciles with
known experiments.
One of important check points is if the predictions coming from inflation in bimetric gravity
are consistent with cosmological observations which are now getting more precise.
We examined the homogeneous expanding solutions and their stability in
the slow-roll limit in our previous paper \cite{Sakakihara:2012iq}.
We studied minimal bimetric models and obtained the unique stable branch of the solutions.
Now, we construct homogeneous inflationary solutions under slow-roll approximation
which correspond to the stable branch.
Then we consider tensor perturbations on the homogeneous solutions and
we calculate primordial tensor spectra generated during inflation
to the first order of the slow-roll parameter.

We note there are works dealing background solutions \cite{vonStrauss:2011mq,Akrami:2012vf}, treating perturbations on the FLRW background \cite{Konnig:2014xva,Lagos:2014lca,Cusin:2014psa,Johnson:2015tfa} and on the detectability of gravitational waves \cite{Corda:2007zz} in bimetric gravity.

This paper is organized as follows. 
In section \ref{sec.IBS}, we introduce an inflaton to a minimal
bimetric model and we construct inflationary background solutions.
We explain the properties of the functions which are specific to bimetric gravity. 
We briefly mention slow-roll approximation and introduce a slow-roll
parameter we use in the following calculation.
In section \ref{sec.SOL}, we derive the second order Lagrangian for
tensor perturbations. There are two views of this system:
the flavor eigenstates and the mass eigenstates, and we use the
mass eigenstates for simplifying calculations.
In section \ref{sec.TS}, we calculate the primordial tensor spectra up
to the first order of the slow-roll parameter by making use of the
interaction picture.
In the final section, we discuss the features of the tensor power-spectra in bimetric theory.

\section{Inflationary background solutions in bimetric gravity}\label{sec.IBS}
In this section, we construct inflationary background solutions with
a scalar field coupled to the physical metric in bimetric gravity. 
We write a minimal bimetric action with a canonical scalar field.
We substitute a homogeneous isotropic metric ansatz into the action 
and derive the equations of motion by using the variational principle. 
We mention the features of the solutions based on our analysis in our
previous paper \cite{Sakakihara:2012iq}.

\subsection{Action and Ansatz}
We consider a minimal bimetric action including a scalar field coupled to the physical metric
and we substitute a homogeneous ansatz for metrics and the scalar field into the action.

We use $g_{\mu\nu}$ as the physical metric, $f_{\mu\nu}$ as the other metric
 and $\varphi$ as the scalar field.
When we consider a bimetric theory \cite{Hassan:2011zd,Hassan:2011ea},
the form of the interaction terms are restricted 
in order to avoid the Boulware-Deser ghost \cite{Boulware:1973my},
which include five theoretical parameters $\{\alpha_n\}(n=0,1,\cdots,4)$.
For simplicity, we set $\alpha_2=1$ and other four theoretical parameters are equal to zero.
Then we have
\begin{eqnarray} 
   S&=&\frac{M_{g}^{2}}{2}\int d^{4}x \sqrt{-g}R[g_{\mu\nu}]
    +\int d^{4}x \sqrt{-g}\Bigl(-\frac{1}{2}g^{\mu\nu}\partial_{\mu}\varphi\partial_{\nu}\varphi-V[\varphi]\Bigr)\nonumber \\
    &&+\frac{M_{f}^{2}}{2}\int d^{4}x \sqrt{-f}R[f_{\mu\nu}]
    +m^{2}M_{e}^{2}\int d^{4}x \sqrt{-g} \,\frac{1}{2}[(K_{\mu}^{\mu})^2-K_{\nu}^{\mu}K_{\mu}^{\nu}] \ ,
\end{eqnarray}
where $M_g$ and $M_f$ are the Planck scales of the physical metric and the
other metric respectively and we defined the reduced Planck scale as 
\begin{eqnarray}
 1/M_{e}^{2}:=1/M_{g}^{2}+1/M_{f}^{2} \ .
\end{eqnarray}
Indices written in Greek letters run over $0, \cdots,3$. 
The last part in the action includes the interaction terms of the physical
metric and the other metric where we defined 
\begin{eqnarray}
 K_{\nu}^{\mu}:= \delta_{\nu}^{\mu}-(\sqrt{g^{-1}f})_{\nu}^{\mu} \ .
\end{eqnarray}

We impose the homogeneous isotropic ansatz for these metrics as
\begin{eqnarray}
	g_{\mu\nu}dx^\mu dx^\nu=-N^{2}(t)dt^{2}+e^{2\alpha(t)}\gamma_{ij}dx^{i}dx^{j} \ ,
\end{eqnarray}
\begin{eqnarray}
	f_{\mu\nu}dx^\mu dx^\nu=-M^{2}(t)dt^{2}+e^{2\beta(t)}\gamma_{ij}dx^{i}dx^{j} \ ,
\end{eqnarray}
where $\gamma_{ij}$ is three-dimensional flat metric and indices
written in Roman letters run over spatial coordinates,
 i.e. $i=1,\cdots, 3$.
$N$ and $M$ are lapse functions and $e^{\alpha}$ and $e^{\beta}$ are scale factors which
depend only on time.
We require the scalar field is also homogeneous
\begin{eqnarray}
 \varphi=\varphi(t) \ .
\end{eqnarray}
When we substitute the ansatz into the Lagrangian, we find
\begin{eqnarray}
 \mathcal{L}&=&
 M_{g}^{2}e^{3\alpha}\biggl[-\frac{3\dot{\alpha}^{2}}{N}\biggr]
 +e^{3\alpha}\biggl[\frac{\dot{\varphi}^2}{2N}-NV(\varphi)\biggr]
 +M_{f}^{2}e^{3\beta}\biggl[-\frac{3\dot{\beta}^{2}}{M}\biggr]\nonumber \\
 &&+m^{2}M_{e}^{2}e^{3\alpha}\bigl[N(6-9\epsilon+3\epsilon^2)+M(-3+3\epsilon)\bigr] \ ,
\end{eqnarray}
where dots denote the time derivative and $\epsilon$ is the ratio of the scale factors of the physical
metric and the other metric, i.e.
\begin{eqnarray}
 \epsilon:=e^{\beta-\alpha} \ .
\end{eqnarray} 
We can see that $\alpha$, $\beta$ and $\varphi$ are dynamical variables
and $N$ and $M$ are non-dynamical variables included in the Lagrangian linearly.

\subsection{Equations of motion}
We derive the equations of motion of the dynamical variables $\alpha$, $\beta$ and
$\varphi$ and two constraints from the variational principle.
We have a relation between the lapse functions
in order for the two constraints to hold during time evolution \cite{Comelli:2011zm}.
We  also describe the behavior of the solutions of the equations.

From the variations of the Lagrangian with respect to the dynamical variables $\alpha$, $\beta$ and
$\varphi$, we obtain
\begin{eqnarray}
 \frac{1}{N}\Bigl(\frac{\dot{\alpha}}{N}\Bigr)^\cdot
=m^2 \cos^2 a\Bigl(\frac{3}{2}-\epsilon\Bigr)\Bigl(\frac{M}{N}-\epsilon\Bigr)-\frac{1}{2M_g^2}\Bigl(\frac{\dot{\varphi}}{N}\Bigr)^2 \ , 
\label{be a}
\end{eqnarray}
\begin{eqnarray}
 \frac{1}{M}\Bigl(\frac{\dot{\beta}}{M}\Bigr)^\cdot
=-m^2 \sin^2 a\Bigl(\frac{3}{2}-\epsilon\Bigr)\Bigl(1-\frac{N\epsilon}{M}\Bigr)\frac{1}{\epsilon^3} \ ,
\label{be b} 
\end{eqnarray}
\begin{eqnarray}
 \frac{1}{N}\Bigl(\frac{\dot{\varphi}}{N}\Bigr)^\cdot+3\frac{\dot{\alpha}}{N}\frac{\dot{\varphi}}{N}+\frac{\mathrm{d}V}{\mathrm{d}\varphi}=0 \ ,
\label{be s}
\end{eqnarray}
respectively, where we defined a new parameter $\tan a:= M_g/M_f$.
Since $N$ and $M$ are non-dynamical and included only
linearly in the Lagrangian, we obtain two constraints from the variations
of the action with respect to them. One of them is 
\begin{eqnarray}
 \Bigl(\frac{\dot{\alpha}}{N}\Bigr)^2=m^2 \cos^2 a(-2+3\epsilon-\epsilon^2)+\frac{1}{3M_g^2}\Bigl[\frac{1}{2}\Bigl(\frac{\dot{\varphi}}{N}\Bigr)^2+V(\varphi)\Bigr] \ ,
\label{be n} 
\end{eqnarray}
which comes from the variation of the action  with respect to $N$. The variation of the action with respect to $M$ yields the other one 
\begin{eqnarray}
 \Bigl(\frac{\dot{\beta}}{M}\Bigr)^2=m^2 \sin^2 a \frac{1-\epsilon}{\epsilon^3} \ .
\label{be m}
\end{eqnarray}
We set the time derivative of eq.(\ref{be n}) is equal to zero so that the
 constraint is satisfied during time evolution. We combine it with 
eq.(\ref{be a}) and obtain the following equation
\begin{eqnarray}
 \Bigl(\frac{3}{2}-\epsilon\Bigr)\Bigl(\dot{\beta}\frac{N\epsilon}{M}-\dot{\alpha}\Bigr)=0
  \ .
\end{eqnarray}
We can obtain the same equation also by using eq.(\ref{be m}) and eq.(\ref{be b}).
When the first factor is equal to zero, 
the solutions are known to be pathological \cite{Gumrukcuoglu:2011zh,Comelli:2012db,DeFelice:2012mx,Gumrukcuoglu:2012aa,Tasinato:2012ze}.
Therefore, we assume the second factor is equal to
zero,
\begin{eqnarray}
 M=\frac{\mathrm{d}\beta}{\mathrm{d}\alpha}\epsilon N=\zeta \epsilon N 
\label{be mn}
\end{eqnarray}
where we defined
\begin{eqnarray}
 \zeta:=\frac{\mathrm{d}\beta}{\mathrm{d}\alpha} \ .
\end{eqnarray}
Since $N$ is a gauge variable, we can set an arbitrary value for $N$.
Substituting this relation into eq.(\ref{be m}) leads to
\begin{eqnarray}
 \Bigl(\frac{\dot{\alpha}}{N}\Bigr)^2= m^2 \sin^2 a
  \frac{1-\epsilon}{\epsilon} \ .
\label{be m 2}
\end{eqnarray}
This is another expression for the Hubble expansion of the physical
metric.
We can see that the Hubble expansion is a function of $\epsilon$.
By equating eq.(\ref{be n}) and eq.(\ref{be m 2}), we obtain an equation
\begin{eqnarray}
 m^2 \sin^2 a \frac{1-\epsilon}{\epsilon}
= m^2 \cos^2
a(-\epsilon^2+3\epsilon-2)+\frac{1}{3M_g^2}\Bigl[\frac{1}{2}\Bigl(\frac{\dot{\varphi}}{N}\Bigr)^2+V(\varphi)\Bigr]
\ .
\label{epsilon}
\end{eqnarray}
This equation says that we find the values of $\epsilon$ 
if we determine the energy density on the physical spacetime.
We have examined the properties of the roots of this equation 
in the slow-roll limit in our previous paper \cite{Sakakihara:2012iq}.
In the slow-roll limit, we neglect $\dot{\varphi}$ contribution 
and this equation reduces to algebraic equation 
with constant coefficients, therefore, the roots are constant. 
This equation basically has three roots since it is a cubic equation.
One of the roots is always negative, therefore it is not appropriate solution
when we take into account the definition of $\epsilon$.
Another is always larger than 1, therefore it is also
inappropriate because the Hubble expansion, which is written as 
eq.(\ref{be m 2}), becomes imaginary.
The other one has the value between 0 and 1, therefore it is the only
adoptable root as de Sitter spacetime.
We have confirmed that the  root satisfies the Higuchi bound
\cite{Higuchi:1986py,Fasiello:2012rw},
which is the stability condition of de Sitter spacetime with a massive
graviton.

Substituting eq.(\ref{be mn}) into eq.(\ref{be a}) and eq.(\ref{be b}), we obtain
the equation determining the value of $\zeta$, 
\begin{eqnarray}
 \zeta = 1 + \frac{(\frac{\dot{\varphi}}{N})^2}{M_g^2\bigl[m_{\rm
 eff}^2(\epsilon)-2(\frac{\dot{\alpha}}{N})^2\bigr]} \ 
\label{delta_z}
\end{eqnarray}
where we defined the effective mass
\begin{eqnarray}
 m_{\rm eff}^2(\epsilon):=m^2\Bigl[\epsilon\cos^2 a
+\frac{1}{\epsilon}\sin^2 a\Bigr](3-2\epsilon) \ .
\label{m eff}
\end{eqnarray}
In the slow-roll limit, $\zeta$ is equal to 1 
and therefore $\alpha$ and $\beta$ are different only by a constant.
Using the definition of $\zeta$, we have a relation between $\zeta$ and
$\epsilon$ as
\begin{eqnarray}
 \delta\zeta := \zeta -1 = \frac{\dot{\epsilon}}{\dot{\alpha}\epsilon} \ .
\label{zeta_eps}
\end{eqnarray}
We can see that $\delta\zeta$ vanishes in the slow-roll limit.

\subsection{Slow-roll approximation}

We introduce a slow-roll parameter 
and explain the assumptions used in the following analysis.
We use the $N=1$ gauge in this subsection.

We define a slow-roll parameter as
\begin{eqnarray}
 s:= -\frac{\dot{H}}{H^2}  \ .
\end{eqnarray}
where $H:=\dot{\alpha}$.
From eq.(\ref{delta_z}), we find
\begin{eqnarray}
 \delta \zeta
=\frac{\dot{\varphi}^2}{M_g^2(m_{\rm eff}^2-2H^2)}
=\frac{2sH^2}{m_{\rm eff}^2-2H^2-m^2(\cos^2 a)\epsilon(3-2\epsilon)}
=2s(1-\epsilon) \ .
\label{zeta}
\end{eqnarray}
The third expression is obtained by eliminating $\dot{\varphi}$ with
eq.(\ref{be a}) and the final expression is obtained by using
eq.(\ref{be m 2}) and eq.(\ref{m eff}).
From this equation, we can read off an expression of $\dot{\varphi}$ as
\begin{eqnarray}
 \dot{\varphi}^2
=2s(1-\epsilon)M_g^2(m_{\rm eff}^2-2H^2)
=2M_g^2(-\dot{H})(1-\epsilon)\Bigl(\frac{m_{\rm eff}^2}{H^2}-2\Bigr)
=2M_g^2(-\dot{H})\Bigl(1+\frac{\epsilon^2(3-2\epsilon)}{\tan^2 a}\Bigr) \ . \nonumber \\
\end{eqnarray}
The third expression is obtained by using the definition of $s$
and the last expression is obtained by using
eq.(\ref{be m 2}) and eq.(\ref{m eff}).
We can see that the expression for $\dot{\varphi}$ is slightly changed
from the conventional case .

We assume the slow-roll parameter is small and constant.
We neglect the higher order contribution $O(s^2)$ in the following.
Under this approximation, we can see $\delta \zeta$ is constant i.e. $\zeta$ is constant but
$\epsilon$ has time dependence.
We can easily obtain the explicit form of $\epsilon$ by integrating 
the differential equation $\dot{\epsilon}=2s\epsilon H (1-\epsilon)$ 
obtained from eq.(\ref{zeta_eps}) and eq.(\ref{zeta}):
\begin{eqnarray}
 \epsilon=\frac{\epsilon_0 e^{2s\alpha}}{(1-\epsilon_0)+\epsilon_0
  e^{2s\alpha}} \ .
\end{eqnarray}
$\epsilon_0$ is the value of $\epsilon$ in the slow-roll limit.
We note that the functions with the subscript $0$ have the values in the slow-roll limit
and all of them are constant in the following.

\section{Second order Lagrangian for tensor perturbations}\label{sec.SOL}
We derive the second order Lagrangian for tensor perturbations
on the homogeneous isotropic inflationary background.
First, we obtain the Lagrangian in the flavor eigenstates,
and then, we move to the mass eigenstates.

\subsection{Flavor eigenstates}

We give tensor perturbations to the original Lagrangian and derive the second
order Lagrangian in the flavor eigenstates.
We will see the variables in the flavor eigenstates are not decoupled
from each other in the slow-roll limit.

We consider perturbations such as
\begin{eqnarray}
 \delta g_{ij} = e^{2\alpha}q_{ij} \ , \quad \delta f_{ij}=e^{2\beta}p_{ij}
\end{eqnarray}
which satisfy the following transverse traceless conditions:
\begin{eqnarray}
 q^{i}{}_{j|i}=0 \ , \quad q^{i}{}_{i}=0 \ , \quad  p^{i}{}_{j|i}=0 \ , \quad p^{i}{}_{i}=0 \ .
\end{eqnarray}
We can decompose the tensor perturbations using polarization tensors as
\begin{eqnarray}
 q_{ij}=q_{I}e^{I}_{ij} \ , \quad  p_{ij}=p_{I}e^{I}_{ij} \ ,
\end{eqnarray}
where the subscripts $I$ correspond to the plus mode and the cross mode,
i.e. $I=\{+,\times\}$.
When we substitute the tensor perturbations into the action,
the second order Lagrangian is reduced to
\begin{eqnarray}
 \delta^2\mathcal{L}
=\sum_{I}&&\frac{M_g^2}{4}Ne^{3\alpha}\Bigl[\frac{1}{2}\frac{\dot{q_I}^2}{N^2}-\frac{1}{2}\frac{k^2}{e^{2\alpha}}q_I^2+\Bigl[-3\Bigl(\frac{\dot{\alpha}}{N}\Bigr)^2-\frac{2}{N}\Bigl(\frac{\dot{\alpha}}{N}\Bigr)^\cdot\Bigr]q_I^2\Bigr]\nonumber \\
&&+\frac{M_f^2}{4}Me^{3\beta}
\Bigl[\frac{1}{2}\frac{\dot{p_I}^2}{M^2}
-\frac{1}{2}\frac{k^2}{e^{2\beta}}p_I^2
+\Bigl[-3\Bigl(\frac{\dot{\beta}}{M}\Bigr)^2-\frac{2}{M}\Bigl(\frac{\dot{\beta}}{M}\Bigr)^\cdot\Bigr]p_I^2\Bigr]\nonumber \\
&&-\frac{1}{4}Ne^{3\alpha}\Bigl[\frac{1}{2}\Bigl(\frac{\dot{\varphi}}{N}\Bigr)^2-V(\varphi)\Bigr]q_I^2\nonumber \\
&&+\frac{1}{4}m^2 M_e^2 e^{3\alpha}\Bigl[\{N(-6+9\epsilon-3\epsilon^2)+M(3-3\epsilon)\}q_I^2
+\{N(6\epsilon-4\epsilon^2)-2M\epsilon\}q_I(p_I-q_I)\nonumber \\
&& \qquad\qquad \quad+\Bigl\{N\Bigl(\frac{3}{2}\epsilon-\frac{3}{2}\epsilon^2\Bigr)-\frac{1}{2}M\epsilon\Bigr\}(p_I-q_I)^2\Bigr] \ ,
\end{eqnarray}
where $k$ is the norm of a three-dimensional wave vector since we have used the Fourier decomposition.
We drop the subscripts $I$ in the following and we finally sum up the two polarization states, or multiply results by two.
Substituting background equations, we obtain
\begin{eqnarray}
 \delta^2\mathcal{L}
&=&\frac{M_g^2}{4}Ne^{3\alpha}\Bigl[\frac{1}{2}\frac{\dot{q}^2}{N^2}-\frac{1}{2}\frac{k^2}{e^{2\alpha}}q^2\Bigl]
+\frac{M_f^2}{4}Me^{3\beta}
\Bigl[\frac{1}{2}\frac{\dot{p}^2}{M^2}
-\frac{1}{2}\frac{k^2}{e^{2\beta}}p^2\Bigr]\nonumber \\
&&+\frac{1}{4}m^2 M_e^2
 e^{3\alpha}\Bigl[N\Bigl(-\frac{3}{2}\epsilon+\frac{1}{2}\epsilon^2\Bigr)+M\frac{1}{2}\epsilon\Bigr](p-q)^2\ .
\end{eqnarray}
Furthermore, if we substitute the consistency relation $M=\zeta\epsilon
N$, this becomes
\begin{eqnarray}
 \delta^2 \mathcal{L}
&=&\frac{Ne^{3\alpha}}{4}\biggl[M_g^2\Bigl(\frac{\dot{q}^2}{2N^2}-\frac{k^2}{2e^{2\alpha}}q^2\Bigr)+M_f^2\epsilon^2\Bigl(\frac{\dot{p}^2}{2\zeta
N^2}-\frac{\zeta k^2}{2e^{2\alpha}}p^2\Bigr) \nonumber\\
&&\qquad
 \qquad-\frac{m^2M_e^2}{2}(3\epsilon-\epsilon^2-\zeta\epsilon^2)(p-q)^2\biggr]
 \ .
\label{Lag_qp}
\end{eqnarray}
The interaction term is proportional to $(p-q)^2$ and do not disappear in
the slow-roll limit where $\epsilon=\epsilon_0$ and $\zeta=1$.
Thus, it is difficult to treat them analytically. 
Hence, we change the variables in the next subsection in order to have decoupled equations 
in the slow-roll limit, which allows us to calculate power-spectra analytically.

\subsection{Mass eigenstates}

We calculate the second order Lagrangian in the mass eigenstates 
by changing the variables in the flavor eigenstates to those in the mass eigenstates.
Then, the calculation becomes much simpler
because a massive eigenstate stemming from the interaction terms 
and a massless eigenstate due to the general covariance
 do not interact with each other in de Sitter spacetime in bimetric gravity \cite{Berg:2012kn}.

We make a transformation of the variables $(q, p)$ into $(x, y)$ as
\begin{eqnarray}
\begin{pmatrix}
 q \\ p
\end{pmatrix}
=
\frac{1}{(\kappa^2+\epsilon^2)^{1/2}}
\begin{pmatrix}
 \kappa & -\epsilon \\ \kappa & \kappa^2/\epsilon
\end{pmatrix}
\begin{pmatrix}
 x \\ y
\end{pmatrix}
\end{eqnarray}
where $\kappa=\zeta^{1/2}\tan a$.
We note that $\kappa$ is constant since $\zeta$ is constant but
$\epsilon$ is time dependent under slow-roll approximation as we mentioned in the last part of the previous
section.
Therefore, this mixing matrix is time dependent.
We can see that $y$ corresponds to a massive mode since $y$ is represented as $p-q$
which combination appears in the mass terms and 
$x$ corresponds to a massless mode since $x$ is orthogonal to $y$.
By substituting this relation into eq.(\ref{Lag_qp}), we find the second order Lagrangian 
in the mass eigenstates up to the first order of the slow-roll parameter as
\begin{eqnarray}
 \delta^2 \mathcal{L}
 &=&\frac{M_g^2}{4}e^{3\alpha}\Bigl[\frac{1}{2}\dot{x}^2+\frac{1}{2}\dot{y}^2
+\frac{3}{2}H^2\frac{\epsilon^2}{\kappa^2+\epsilon^2}\delta \zeta x^2
+\frac{3}{2}H^2\frac{\kappa^2}{\kappa^2+\epsilon^2}\delta \zeta y^2
-2H\frac{\kappa\epsilon}{\kappa^2+\epsilon^2}\delta \zeta\dot{x}y 
-\frac{1}{2}\tilde{m}_{\rm eff}^2 y^2 \nonumber \\
&& \qquad -\frac{1}{2}\frac{k^2}{e^{2\alpha}}\Bigl(1+\frac{2\epsilon^2}{\kappa^2+\epsilon^2}\delta\zeta\Bigr)x^2
-2\frac{k^2}{e^{2\alpha}}\frac{\kappa\epsilon}{\kappa^2+\epsilon^2}\delta\zeta xy
 -\frac{1}{2}\frac{k^2}{e^{2\alpha}}\Bigl(1+\frac{2\kappa^2}{\kappa^2+\epsilon^2}\delta\zeta\Bigr)y^2\Bigr]
\label{Lag_xy}
\end{eqnarray}
in the $N=1$ gauge, where
\begin{eqnarray}
 \tilde{m}_{\rm eff}^2
&:=&m^2 \cos^2 a \frac{3-\epsilon-\zeta\epsilon}{\epsilon}(\kappa^2+\epsilon^2)\nonumber\\
&=&m^2\biggl[\epsilon\cos^2 a+\frac{\zeta}{\epsilon}\sin^2 a\biggr](3-\epsilon-\zeta\epsilon) \nonumber \\
&=&m_{\rm eff}^2(\epsilon)+2s(1-\epsilon_0)[3H_0^2-m^2\epsilon_0^2\cos^2
 a] 
\label{m_eff_tilde}
\end{eqnarray}
and
\begin{eqnarray}
 H_0^2:=m^2 \sin^2 a \frac{1-\epsilon_0}{\epsilon_0} \ .
\end{eqnarray}
We have used the relation $\dot{\epsilon} = \epsilon H\delta\zeta$ 
in eq.(\ref{Lag_xy}) and also eq.(\ref{zeta}) in the last line of eq.(\ref{m_eff_tilde}).
If we use the $N=e^\alpha$ gauge, this Lagrangian is rewritten as
\begin{eqnarray}
 \delta^2 \mathcal{L}
&=&\frac{M_g^2}{4}e^{2\alpha}\Bigl[\frac{1}{2}x^{\prime 2}+\frac{1}{2}y^{\prime 2}
+\frac{3}{2}\mathcal{H}^2\frac{\epsilon^2}{\kappa^2+\epsilon^2}\delta \zeta x^2
+\frac{3}{2}\mathcal{H}^2\frac{\kappa^2}{\kappa^2+\epsilon^2}\delta \zeta y^2
-2\mathcal{H}\frac{\kappa\epsilon}{\kappa^2+\epsilon^2}\delta \zeta x^{\prime}y 
-\frac{1}{2}\tilde{m}_{\rm eff}^2e^{2\alpha} y^2 \nonumber \\
&& \qquad -\frac{1}{2}k^2\Bigl(1+\frac{2\epsilon^2}{\kappa^2+\epsilon^2}\delta\zeta\Bigr)x^2
-2k^2\frac{\kappa\epsilon}{\kappa^2+\epsilon^2}\delta\zeta xy
 -\frac{1}{2}k^2\Bigl(1+\frac{2\kappa^2}{\kappa^2+\epsilon^2}\delta\zeta\Bigr)y^2
\Bigr]   \ .
\end{eqnarray}
The primes denotes the time derivative with respect to the conformal time
defined as $dt=e^{\alpha}d\eta$ and $\mathcal{H}:=\alpha^{\prime}$.
Furthermore, we make a scale transformation to make kinetic terms canonical as
\begin{eqnarray}
\begin{pmatrix}
 x \\ y
\end{pmatrix}
=
\frac{2}{M_ge^{\alpha}}
\begin{pmatrix}
 X \\ Y
\end{pmatrix}\ .
\end{eqnarray}
Then, we obtain the following Lagrangian.
\begin{eqnarray}
 \delta^2 \mathcal{L}
&=&\frac{1}{2}X^{\prime 2}
+\frac{1}{2}\Bigl[\Bigl(2-s+\frac{3\epsilon^2}{\kappa^2+\epsilon^2}\delta\zeta\Bigr)\mathcal{H}^2
-k^2\Bigl(1+\frac{2\epsilon^2}{\kappa^2+\epsilon^2}\delta\zeta\Bigr)\Bigr]X^2 \nonumber \\
&&+\frac{1}{2}Y^{\prime 2}
+\frac{1}{2}\Bigl[\Bigl(2-s-\frac{\tilde{m}_{\rm eff}^2}{H^2}
+\frac{3\kappa^2}{\kappa^2+\epsilon^2}\delta\zeta\Bigr)\mathcal{H}^2
-k^2\Bigl(1+\frac{2\kappa^2}{\kappa^2+\epsilon^2}\delta\zeta\Bigr)\Bigr]Y^2 \nonumber \\
&&+2\frac{\kappa\epsilon}{\kappa^2+\epsilon^2}\delta \zeta 
(-\mathcal{H}X^{\prime}Y+\mathcal{H}^2XY-k^2XY) \nonumber \\
&=&\frac{1}{2}X^{\prime 2}
+\frac{1}{2}\Bigl[\frac{2+3s+\frac{3\epsilon^2}{\kappa^2+\epsilon^2}\delta\zeta}{\eta^2}
-k^2\Bigl(1+\frac{2\epsilon^2}{\kappa^2+\epsilon^2}\delta\zeta\Bigr)\Bigr]X^2 \nonumber \\
&&+\frac{1}{2}Y^{\prime 2}
+\frac{1}{2}\Bigl[\frac{2+3s-\frac{(1+2s)\tilde{m}_{\rm eff}^2}{H^2}
+\frac{3\kappa^2}{\kappa^2+\epsilon^2}\delta\zeta}{\eta^2}
-k^2\Bigl(1+\frac{2\kappa^2}{\kappa^2+\epsilon^2}\delta\zeta\Bigr)\Bigr]Y^2 \nonumber \\
&&+\frac{2\kappa\epsilon}{\kappa^2+\epsilon^2}\delta \zeta 
\Bigl(\frac{1}{\eta}X^{\prime}Y+\frac{1}{\eta^2}XY-k^2XY\Bigr)  \ .
\end{eqnarray}
We have used the relations $\mathcal{H}^{\prime}=(1-s)\mathcal{H}$ and 
$H=\mathcal{H}/e^{\alpha}$ in the first line and 
$\mathcal{H}=-1/(1-s)\eta$ in the second line.
By substituting eq.(\ref{zeta}) and neglecting $\mathcal{O}(s^2)$, we
finally obtain
\begin{eqnarray}
 \delta^2 \mathcal{L}
&=&\frac{1}{2}X^{\prime 2}
+\frac{1}{2}\Biggl[\frac{2+3s\bigl(1+\frac{2\epsilon_0^2(1-\epsilon_0)}{\kappa_0^2+\epsilon_0^2}\bigr)}{\eta^2}-k^2\Bigl(1+\frac{4s\epsilon_0^2(1-\epsilon_0)}{\kappa_0^2+\epsilon_0^2}\Bigr)\Biggr]X^2\nonumber \\
&&+\frac{1}{2}Y^{\prime 2}
+\frac{1}{2}\Biggl[\frac{2+3s\bigl(1+\frac{2\kappa_0^2(1-\epsilon_0)}{\kappa_0^2+\epsilon_0^2}\bigr)-\frac{(1+2s)\tilde{m}_{\rm eff}^2}{H^2}}{\eta^2}-k^2\Bigl(1+\frac{4s\kappa_0^2(1-\epsilon_0)}{\kappa_0^2+\epsilon_0^2}\Bigr)\Biggr]Y^2 \nonumber \\
&&+s\frac{4\kappa_0\epsilon_0(1-\epsilon_0)}{\kappa_0^2+\epsilon_0^2}\Bigl(\frac{1}{\eta}X'Y+\frac{1}{\eta^2}XY-k^2XY\Bigr)  
\end{eqnarray}
where $\kappa_0=\tan a$.
We note that the cross terms of $X$ and $Y$ vanish in the slow-roll
limit and the Lagrangian becomes diagonal in the mass eigenstates.
We can see that the propagation speed of the massless graviton differs from that of the conventional case 
under slow-roll approximation due to the difference between the lapse function of the physical metric 
and that of the other metric.
The coefficients of the terms proportional to $1/\eta^2$ also differ from that of the conventional case
as the result of time dependence of the mixing matrix, 
which causes the modification of the wavenumber dependence of tensor power-spectra.

Since we have obtained the second order Lagrangian of tensor perturbations in the mass eigenstates, 
in which the modes are decoupled from each other in the slow-roll limit,
we are now ready to calculate the tensor power-spectra.

\section{Tensor power-spectra in bimetric gravity}\label{sec.TS}

We calculate the tensor power-spectra generated during inflation in bimetric gravity.
In the mass eigenstates, we can solve the free part. Then, we calculate the first order
corrections for them by using the interaction picture 
and obtain tensor power-spectra in the mass eigenstates.
Finally, we obtain the tensor power-spectra in the flavor eigenstates
by using the relation between the variables in the
flavor eigenstates and those in the mass eigenstates.

\subsection{Interaction picture}
First, we transform the Lagrangian to the Hamiltonian and quantize the system. 
Then, we calculate the correlations of the massless state and
the massive state by making use of the interaction picture.

The conjugate momenta of $X$ and $Y$ are calculated as
\begin{eqnarray}
 \pi_X = \frac{\partial {\mathcal L}}{\partial X'}=X'+\tilde{s}\frac{Y}{\eta} \ , \qquad 
 \pi_Y = \frac{\partial {\mathcal L}}{\partial Y'}=Y'\ ,
\end{eqnarray}
where we defined
\begin{eqnarray}
 \tilde{s}=s\frac{4\kappa_0\epsilon_0(1-\epsilon_0)}{\kappa_0^2+\epsilon_0^2} \ .
\end{eqnarray}
By performing Legendre transformations, we obtain the following Hamiltonian.
\begin{eqnarray}
 {\mathcal H}&=&\pi_X X'+\pi_Y Y'-\delta^2 {\mathcal L} \nonumber \\
 &=&\frac{1}{2}\pi_X^{2}
+\frac{1}{2}\Biggl[k^2\Bigl(1+\frac{4s\epsilon_0^2(1-\epsilon_0)}{\kappa_0^2+\epsilon_0^2}\Bigr)-\frac{2+3s\bigl(1+\frac{2\epsilon_0^2(1-\epsilon_0)}{\kappa_0^2+\epsilon_0^2}\bigr)}{\eta^2}\Biggr]X^2\nonumber \\
&&+\frac{1}{2}\pi_Y^{2}
+\frac{1}{2}\Biggl[k^2\Bigl(1+\frac{4s\kappa_0^2(1-\epsilon_0)}{\kappa_0^2+\epsilon_0^2}\Bigr)-\frac{2+3s\bigl(1+\frac{2\kappa_0^2(1-\epsilon_0)}{\kappa_0^2+\epsilon_0^2}\bigr)-\frac{(1+2s)\tilde{m}_{\rm eff}^2}{H^2}}{\eta^2}\Biggr]Y^2 \nonumber \\
&&+\tilde{s}\Bigl(-\frac{1}{\eta}\pi_X Y-\frac{1}{\eta^2}XY+k^2XY\Bigr)  \ .
\end{eqnarray}
We note that we have defined $X({\bf k}, \eta)$ and $Y({\bf k}, \eta)$ as 
\begin{eqnarray}
 X(x)=\int \frac{d^3 k}{(2\pi)^{3/2}}e^{i{\bf k}\cdot{\bf x}}X({\bf
  k}, \eta) \ ,
 \qquad  Y(x)=\int \frac{d^3 k}{(2\pi)^{3/2}}e^{i{\bf k}\cdot{\bf x}}Y({\bf
  k}, \eta) \ .
\end{eqnarray}
We define free fields $X_I$ and $Y_I$ where the time evolution is governed by the Hamiltonian
\begin{eqnarray}
 H_0
 &:=&\int d^3 k\;\frac{1}{2}\pi_{X_I}^2
+\frac{1}{2}\Biggl[k^2\Bigl(1+\frac{4s\epsilon_0^2(1-\epsilon_0)}{\kappa_0^2+\epsilon_0^2}\Bigr)-\frac{2+3s\bigl(1+\frac{2\epsilon_0^2(1-\epsilon_0)}{\kappa_0^2+\epsilon_0^2}\bigr)}{\eta^2}\Biggr]X_I^2\nonumber \\
&&+\frac{1}{2}\pi_{Y_I}^2
+\frac{1}{2}\Biggl[k^2\Bigl(1+\frac{4s\kappa_0^2(1-\epsilon_0)}{\kappa_0^2+\epsilon_0^2}\Bigr)-\frac{2+3s\bigl(1+\frac{2\kappa_0^2(1-\epsilon_0)}{\kappa_0^2+\epsilon_0^2}\bigr)-\frac{(1+2s)\tilde{m}_{\rm eff}^2}{H^2}}{\eta^2}\Biggr]Y_I^2  \ .
\end{eqnarray}
We separated the interaction Hamiltonian from the free one
\begin{eqnarray}
 H_{\rm int}:= \int d^3k\;\tilde{s}\Bigl(-\frac{1}{\eta}\pi_{X_I}
  Y_I-\frac{1}{\eta^2}X_I Y_I+k^2X_I Y_I\Bigr) \ .
\end{eqnarray}
The fields $X_I$ and $Y_I$ can be expanded by creation and annihilation operators 
\begin{eqnarray}
 X_I({\bf k}, \eta)
=u_X(k, \eta)a_{\bf k}+u_X^{\ast}(k,  \eta)a_{\bf -k}^{\dagger} \ ,
\end{eqnarray}
\begin{eqnarray}
 Y_I({\bf k}, \eta)=u_Y(k, \eta)b_{\bf k}+u_Y^{\ast}(k,
  \eta)b_{\bf -k}^{\dagger} \ ,
\end{eqnarray}
where mode functions satisfy
\begin{eqnarray}
 u_X''+\Biggl[k^2\Bigl(1+\frac{4s\epsilon_0^2(1-\epsilon_0)}{\kappa_0^2+\epsilon_0^2}\Bigr)-\frac{2+3s\bigl(1+\frac{2\epsilon_0^2(1-\epsilon_0)}{\kappa_0^2+\epsilon_0^2}\bigr)}{\eta^2}\Biggr]u_X
=0 \ ,
\label{eq.ux}
\end{eqnarray}
and
\begin{eqnarray}
 u_Y''+\biggl[k^2\Bigl(1+\frac{4s\kappa_0^2(1-\epsilon_0)}{\kappa_0^2+\epsilon_0^2}\Bigr)-\frac{2+3s\bigl(1+\frac{2\kappa_0^2(1-\epsilon_0)}{\kappa_0^2+\epsilon_0^2}\bigr)-\tilde{m}_{\rm eff}^2/H^2}{\eta^2}\biggr]u_Y
=0 \ .
\label{eq.uy}
\end{eqnarray}
The correlation functions of these variables can be deduced as
\begin{eqnarray}
 \langle X_I({\bf k}, \eta)X_I({\bf k}^{\prime}, \eta^{\prime})\rangle
 =u_X(k,\eta)u_X^\ast(k,\eta^\prime)\delta^3({\bf k}+{\bf k}^\prime) \ ,
\end{eqnarray}
\begin{eqnarray}
 \langle X_I({\bf k}, \eta)Y_I({\bf k}^{\prime}, \eta^{\prime})\rangle
 =0 \ ,
\end{eqnarray}
and
\begin{eqnarray}
 \langle Y_I({\bf k}, \eta)Y_I({\bf k}^{\prime}, \eta^{\prime})\rangle
 =u_Y(k,\eta)u_Y^\ast(k,\eta^\prime)\delta^3({\bf k}+{\bf k}^\prime) \ .
\end{eqnarray}
The correlation functions of the original variables are calculated by
using the interaction picture, up to
the first order of the slow-roll parameter, as
\begin{eqnarray}
 &&\langle X({\bf k},\eta)X({\bf k}^\prime,\eta) \rangle \nonumber \\
&=&\langle X_I({\bf k},\eta) X_I({\bf k}^\prime,\eta) \rangle
 + i \int_{-\infty}^\eta d\eta_1 (\langle  H_{\rm int}(\eta_1) X_I({\bf
 k},\eta)X_I({\bf k}^\prime,\eta)\rangle-\langle X_I({\bf
 k},\eta)X_I({\bf k}^\prime,\eta) H_{\rm int}(\eta_1)\rangle)  \nonumber \\
&=&\langle X_I({\bf k},\eta) X_I({\bf k}^\prime,\eta) \rangle
=\left|u_X(\eta)\right|^2 \delta^3({\bf k}+{\bf k}^\prime)
\end{eqnarray}
and
\begin{eqnarray}
 &&\langle Y({\bf k},\eta)Y({\bf k}^\prime,\eta) \rangle \nonumber \\
&=&\langle Y_I({\bf k},\eta) Y_I({\bf k}^\prime,\eta) \rangle
 + i \int_{-\infty}^\eta d\eta_1 (\langle  H_{\rm int}(\eta_1) Y_I({\bf
 k},\eta)Y_I({\bf k}^\prime,\eta)\rangle-\langle Y_I({\bf
 k},\eta)Y_I({\bf k}^\prime,\eta) H_{\rm int}(\eta_1)\rangle)  \nonumber \\
&=&\langle Y_I({\bf k},\eta) Y_I({\bf k}^\prime,\eta) \rangle
=\left|u_Y(\eta)\right|^2 \delta^3({\bf k}+{\bf k}^\prime)\ .
\end{eqnarray}
In the lowest order, there is no effect of interactions.
However, the cross correlation gets corrections as
\begin{eqnarray}
&& \langle X({\bf k},\eta)Y({\bf k}^\prime,\eta) \rangle \nonumber \\
&=&\langle X_I({\bf k},\eta) Y_I({\bf k}^\prime,\eta) \rangle
 + i \int_{-\infty}^\eta d\eta_1 (\langle  H_{\rm int}(\eta_1) X_I({\bf k},\eta)Y_I({\bf k}^\prime,\eta)\rangle-\langle X_I({\bf k},\eta)Y_I({\bf k}^\prime,\eta) H_{\rm int}(\eta_1)\rangle)  \nonumber \\
&=& -2\Im \int_{-\infty}^\eta d\eta_1 \langle H_{\rm int}(\eta_1) X_I({\bf k},\eta)Y_I({\bf k}^\prime,\eta) \rangle \nonumber \\
&=& 2\tilde{s}\Im \int_{-\infty}^\eta d\eta_1\int d^3k_1 \;
 \frac{1}{\eta_1}\langle \frac{d}{d\eta_1}X_I({\bf
 k}_1, \eta_1)Y_I(-{\bf k}_1,\eta_1)X_I({\bf k}, \eta) Y_I({\bf
 k}^\prime, \eta) \rangle \nonumber \\
&&\qquad \qquad \qquad  \qquad \qquad
+\Bigl(\frac{1}{\eta^2_1}-k_1^2\Bigr) \langle X_I({\bf k}_1,\eta_1)Y_I(-{\bf k}_1,\eta_1)X_I({\bf k},\eta)Y_I({\bf k}^\prime,\eta)\rangle \nonumber \\
&=& 2\tilde{s}\delta^3({\bf k}+{\bf k^\prime})\Im \int_{-\infty}^\eta
 d\eta_1 \;
 \frac{1}{\eta_1}u_X^\prime(k,\eta_1)u_X^\ast(k,\eta)u_Y(k,\eta_1)u_Y^\ast(k,\eta)
\nonumber \\
 &&\qquad \qquad \qquad  \qquad \qquad
+\Bigl(\frac{1}{\eta^2_1}-k^2\Bigr)u_X(k,\eta_1)u_X^\ast(k,\eta)u_Y(k,\eta_1)u_Y^\ast(k,\eta)
\ .
\end{eqnarray}

We will calculate these correlations explicitly in the long wavelength 
limit in the following.

\subsection{Power-spectra in the mass eigenstates}
We calculate the mode functions up to the order needed for calculation 
and obtain the tensor power-spectra in the mass eigenstates.
We will see that the amplitude of the power-spectrum of the massless state remains at the
end of inflation, while those of the cross spectrum and the
power-spectrum of the massive state rapidly decay and are negligible.

\subsubsection{Power-spectrum of the massless state}

First, we calculate the power-spectrum of the massless state.  We see
that the power-spectrum looks like that of general relativity.

In the remote past $\eta\rightarrow -\infty$, the differential equations of motion for the mode functions become
\begin{eqnarray}
 u_X^{\prime\prime}+k_X^2u_X=0  \ , \qquad 
 u_Y^{\prime\prime}+k_Y^2u_Y=0 \ ,
\end{eqnarray}
 where
\begin{eqnarray}
 k_X=k\Bigl(1+\frac{4s\epsilon_0^2(1-\epsilon_0)}{\kappa_0^2+\epsilon_0^2}\Bigr)^{1/2}
 \ , \qquad 
 k_Y=k\Bigl(1+\frac{4s\kappa_0^2(1-\epsilon_0)}{\kappa_0^2+\epsilon_0^2}\Bigr)^{1/2} \ .
\end{eqnarray}
Thus, we adopt the Bunch-Davies vacuum states as initial conditions
\begin{eqnarray}
 u_X(k,\eta) = \frac{1}{\sqrt{2k_X}}e^{-ik_X\eta}
 \ , \qquad
 u_Y(k,\eta) = \frac{1}{\sqrt{2k_Y}}e^{-ik_Y\eta} \ .
\end{eqnarray}
Then, we obtain the solution of eq.(\ref{eq.ux}) as
\begin{eqnarray}
 u_X(k,\eta)
&=&\frac{\sqrt{-\pi\eta}}{2}e^{i\frac{2\nu_X+1}{4}\pi}H_{\nu_X}^{(1)}(-k_X\eta) \nonumber \\
&=&\frac{\sqrt{-\pi\eta}}{2}e^{i\frac{2\nu_X+1}{4}\pi}\frac{i}{\sin
 \nu_X
 \pi}\sum_{n=0}^{\infty}\Bigl[e^{-i\nu_X\pi}\frac{(-1)^n(-k_X\eta/2)^{2n+\nu_X}}{\Gamma(n+1+\nu_X)}-\frac{(-1)^n(-k_X\eta/2)^{2n-\nu_X}}{\Gamma(n+1-\nu_X)}\Bigr]
 \ , \nonumber \\
\end{eqnarray}
where
\begin{eqnarray}
\nu_X = \frac{3}{2}+s\Bigl(1+\frac{2\epsilon_0^2(1-\epsilon_0)}{\kappa_0^2+\epsilon_0^2}\Bigr) \ .
\end{eqnarray}
The Hankel function $H_{\nu_X}^{(1)}(x)$ is defined by
\begin{eqnarray}
 H_{\nu}^{(1)}(z):=\frac{i}{\sin \nu \pi}\sum_{n=0}^{\infty}\Bigl[e^{-i\nu\pi}\frac{(-1)^n(z/2)^{2n+\nu}}{\Gamma(n+1+\nu)}-\frac{(-1)^n(z/2)^{2n-\nu}}{\Gamma(n+1-\nu)}\Bigr] \ .
\end{eqnarray}
This solution reduces to
\begin{eqnarray}
 u_X(k,\eta) \rightarrow \frac{\sqrt{-\pi\eta}}{2}e^{i\frac{2\nu_X+1}{4}\pi}\frac{-i}{\pi}\Gamma(\nu_X)\Bigl(\frac{-k_X\eta}{2}\Bigr)^{-\nu_X}
\end{eqnarray}
for $\eta \sim 0$.
In the large scale limit, the power-spectrum of the massless state is given by
\begin{eqnarray}
 \langle X(z)X(z) \rangle
&=& \int \frac{d^3k d^3 k^\prime}{(2\pi)^3}e^{i({\bf k}+{\bf
k}^\prime)\cdot {\bf z}}\langle X({\bf k},\eta)X({\bf k}^\prime,
\eta)\rangle\nonumber \\
&=& \int \frac{dk}{k}\frac{4\pi k^3}{(2\pi)^3}\left|u_X(k,\eta)\right|^2
 \nonumber \\
&=& \int \frac{dk}{k}\frac{4\pi k^3}{(2\pi)^3}\frac{-\pi\eta}{4}\frac{\Gamma(\nu_X)^2}{\pi^2}\Bigl(\frac{-k_X\eta}{2}\Bigr)^{-2\nu_X}
\end{eqnarray}
where $z^0=\eta$, $z^i=({\bf z})^i$.
Therefore, we obtain
\begin{eqnarray}
 \langle x(z)x(z) \rangle 
&=&\frac{4}{M_g^2e^{2\alpha}} \langle X(z)X(z) \rangle \nonumber \\
&=& \int d(\log k) \biggl(\frac{H_0}{\pi
M_g}\biggr)^2(-H_0\eta)^{2s}\biggl(\frac{-k\eta}{2}\biggr)^{3-2\nu_X}\biggl(\frac{\Gamma(\nu_X)}{\Gamma(\frac{3}{2})}\biggr)^2\Delta^{-2\nu_X}
\end{eqnarray}
where
\begin{eqnarray}
 \Delta=\Bigl(1+\frac{4s\epsilon_0^2(1-\epsilon_0)}{\kappa_0^2+\epsilon_0^2}\Bigr)^{1/2} 
\end{eqnarray}
and $H_0$ is the constant Hubble in the slow-roll limit.
We note that we have used the relation $e^\alpha=(-H_0\eta)^{-1-s}$ obtained from the
definition of $s$ and $\Delta$ comes from the modification of the propagation speed.
This amplitude is almost constant as in the conventional case, i.e. 
$\langle xx \rangle \sim \mathcal{O}(\eta^0)$ .

\subsubsection{Cross power-spectrum}

Next, we calculate the cross spectrum of the massless state and the
massive state. We conclude the cross spectrum is negligible compared with
the power-spectrum of the massless state.

For calculating the cross spectrum in the first order of the slow-roll parameter,
we only have to know the mode functions in the lowest order.
The mode functions of $X$ which we have calculated reduces to
\begin{eqnarray}
 u_X(k,\eta)=\frac{1}{\sqrt{2k}}\Bigl(1-\frac{i}{k\eta}\Bigr)e^{-ik\eta}
\end{eqnarray}
in the slow-roll limit and the solution of eq.(\ref{eq.uy}) is written as
\begin{eqnarray}
 u_Y(k, \eta)
&=&\frac{\sqrt{-\pi\eta}}{2}e^{-\frac{\mu\pi}{2}+i\frac{\pi}{4}}H_{i\mu}^{(1)}(-k\eta)\nonumber \\
&=&\frac{\sqrt{-\pi\eta}}{2}e^{-\frac{\mu\pi}{2}+i\frac{\pi}{4}}\frac{1}{\sinh \mu\pi}\sum_{n=0}^{\infty}\Bigl[e^{\mu\pi}\frac{(-1)^n(-k\eta/2)^{2n+i\mu}}{\Gamma(n+1+i\mu)}-\frac{(-1)^n(-k\eta/2)^{2n-i\mu}}{\Gamma(n+1-i\mu)}\Bigr]\nonumber \\
&=&\frac{\sqrt{-\pi\eta}}{2}e^{i\frac{\pi}{4}}\frac{1}{\sinh \mu\pi}\sum_{n=0}^{\infty}\Bigl[e^{\frac{\mu\pi}{2}}\frac{(-1)^n(-k\eta/2)^{2n+i\mu}}{\Gamma(n+1+i\mu)}-e^{-\frac{\mu\pi}{2}}\frac{(-1)^n(-k\eta/2)^{2n-i\mu}}{\Gamma(n+1-i\mu)}\Bigr]\nonumber \\
\end{eqnarray}
where
\begin{eqnarray}
 \mu=\sqrt{\Bigl(\frac{m_{\rm eff}^2(\epsilon_0)}{H_0^2}-3\Bigr)+\frac{3}{4}} \ .
\end{eqnarray}
Note that $\mu$ is a real number since we have proven $m_{\rm eff}^2(\epsilon_0)> 3H_0^2$ in our previous paper \cite{Sakakihara:2012iq}. 
The cross spectrum of the massless state and the massive state is
\begin{eqnarray}
 &&\langle X({\bf k},\eta)Y({\bf k}^\prime,\eta) \rangle \nonumber \\
&=&2\tilde{s}\delta^3({\bf k}+{\bf k}^\prime)\Im\int_{-\infty}^{\eta}
d\eta_1\biggl[\frac{1}{\eta_1}u_X^\prime({\bf
k},\eta_1)+\Bigl(\frac{1}{\eta_1^2}-k^2\Bigr)u_X({\bf
k},\eta_1)\biggr]u_X^\ast({\bf k},\eta)u_Y({\bf k},\eta_1)u_Y^\ast({\bf
k},\eta)\nonumber \\
&=&2\tilde{s}\delta^3({\bf k}+{\bf k}^\prime)\Im\int_{-\infty}^{\eta}d\eta_1
\frac{-k^2}{\sqrt{2k}}e^{-ik\eta_1}u_X^\ast({\bf k},\eta)u_Y({\bf k},\eta_1)u_Y^\ast({\bf
k},\eta)\nonumber \\
&=&\tilde{s}\delta^3({\bf k}+{\bf k}^\prime)\Im\;\frac{-\pi}{4k}e^{-\pi\mu}
\Bigl(1+\frac{i}{k\eta}\Bigr)e^{ik\eta}\sqrt{-k\eta}H_{i\mu}^{(1)\ast}(-k\eta)
\int_{-k\eta}^\infty dz \; e^{iz}z^{1/2}H_{i\mu}^{(1)}(z)\nonumber \\
&=&\frac{\tilde{s}\eta}{4\pi}\delta^3({\bf k}+{\bf k}^\prime)
\Re \biggl[\biggl\{e^{\frac{\pi\mu}{2}}\Bigl(\frac{-k\eta}{2}\Bigr)^{-i\mu}\Gamma(i\mu)
+e^{-\frac{\pi\mu}{2}}\Bigl(\frac{-k\eta}{2}\Bigr)^{i\mu}\Gamma(-i\mu)\biggr\}\nonumber\\
&& \qquad \qquad \qquad \qquad\times
\biggl\{e^{\frac{\pi\mu}{2}}\Bigl(\frac{-k\eta}{2}\Bigr)^{i\mu}\frac{\Gamma(-i\mu)}{\bigl(\frac{3}{2}+i\mu\bigr)}+e^{-\frac{\pi\mu}{2}}\Bigl(\frac{-k\eta}{2}\Bigr)^{-i\mu}\frac{\Gamma(i\mu)}{\bigl(\frac{3}{2}-i\mu\bigr)}\biggr\}\biggr]+\mathcal{O}(\eta^2)\nonumber \\
&=&\frac{\tilde{s}\eta}{4\pi (m_{\rm eff}^2/H^2)}\delta^3({\bf k}+{\bf k}^\prime)
\biggl[\frac{3\pi}{\mu\tanh \pi\mu}+2\Re
\biggl\{\Bigl(\frac{-k\eta}{2}\Bigr)^{-2i\mu}\Bigl(\frac{3}{2}+i\mu\Bigr)\Gamma^2(i\mu)\biggr\}\biggr]+\mathcal{O}(\eta^2)
\ ,
\end{eqnarray}
and $\langle X(z) Y(z) \rangle$ scales as $\eta$ in the leading order.
Therefore, 
\begin{eqnarray}
 \langle x(z)y(z) \rangle
= \frac{4}{M_g^2e^{2\alpha}}\langle X(z) Y(z) \rangle 
={\mathcal O}(\eta^3)\ .
\end{eqnarray}
Similarly, $\langle y(z)x(z) \rangle=(\langle x(z)y(z) \rangle)^\ast=\mathcal{O}(\eta^3)$.
The amplitude decays as $e^{-3\alpha}$ during inflation and we can
neglect this amplitude
compared with that of the power-spectrum of the massless state.

\subsubsection{Power-spectrum of the massive state}

Finally, we calculate the power-spectrum of the massive state. It turns out that it can also be
neglected compared with the power-spectrum of the massless state.

The mode functions of $Y$ satisfy
\begin{eqnarray}
 u_Y''+\biggl[k_Y^2+\frac{C(\eta)}{\eta^2}\biggr]u_Y=0 \ ,
\label{eq.uy_mod}
\end{eqnarray}
where
\begin{eqnarray}
 C(\eta):=\frac{\tilde{m}_{\rm eff}^2(\epsilon)}{H^2}-2-3s\Bigl(1+\frac{2\kappa_0^2(1-\epsilon_0)}{\kappa_0^2+\epsilon_0^2}\Bigr)
=C_0+s C_1(\eta) \ .
\end{eqnarray}
Here, $C_0$ is defined as $C_0:=m_{\rm eff}^2(\epsilon_0)/H_0^2-2$ and $C_1(\eta)$ is a
function of $\eta$ since $\epsilon$ has time dependence
$\epsilon=\epsilon(\eta)$
under slow-roll approximation.
We define $y=\log(-\eta)$, then, this equation becomes
\begin{eqnarray}
 \frac{d^2}{d^2y}u_Y-\frac{d}{dy}u_Y+(k_Y^2e^{2y}+C(y))u_Y=0 \ .
\end{eqnarray}
From the definition of $y$, we can see $y \rightarrow -\infty$ as $\eta \rightarrow -0$ .
If we decompose the solutions like $u_Y=e^{y/2}f(y)$, then we obtain
\begin{eqnarray}
 \frac{d^2}{d^2 y}f+\Bigl(-\frac{1}{4}+k_Y^2e^{2y}+C(y)\Bigr)f=0 \ .
\end{eqnarray}
We can neglect the term which is proportional to $e^{2y}$ because we are
interested in the behavior of solutions around $\left| \eta \right| \sim 0$.
We obtain WKB solutions of this equation as
\begin{eqnarray}
 f=\frac{1}{(4C-1)^{1/4}}\exp{\biggl[-i\int^{y}dy\Bigl(C-\frac{1}{4}\Bigr)^{1/2}\biggr]}
\end{eqnarray}since
\begin{eqnarray}
 \frac{dC(y)}{dy}={\mathcal O}(s) \ , \quad 
 \frac{d^2C(y)}{d^2y}={\mathcal O}(s^2) \ .
\end{eqnarray}
Finally, we obtain 
\begin{eqnarray}
 u_Y=\frac{e^{\frac{y}{2}}}{(4C-1)^{1/4}}\exp{\biggl[-i\int^{y}dy\Bigl(C-\frac{1}{4}\Bigr)^{1/2}\biggr]}
\end{eqnarray}
up to the overall coefficient.
From this, we can calculate the power-spectrum of the massive state
\begin{eqnarray}
 \langle Y(z)Y(z) \rangle
&=& \int \frac{d^3k d^3 k^\prime}{(2\pi)^3}e^{i({\bf k}+{\bf
k}^\prime)\cdot {\bf z}}\langle Y({\bf k},\eta)Y({\bf k}^\prime,
\eta)\rangle\nonumber \\
&=& \int \frac{dk}{k}\frac{4\pi k^3}{(2\pi)^3}\left|u_Y(k,\eta)\right|^2
 \nonumber \\
&=& \int \frac{dk}{k}\frac{4\pi
 k^3}{(2\pi)^3}\frac{-\eta}{(4C-1)^{1/2}} = {\mathcal O}(\eta) \ .
\end{eqnarray}
Thus, we reach the final result
\begin{eqnarray}
 \langle y(z)y(z) \rangle
= \frac{4}{M_g^2e^{2\alpha}}\langle Y(z) Y(z) \rangle
= {\mathcal O} (\eta^3)\ .
\end{eqnarray}
As mentioned before, we can neglect this amplitude compared with that of the power-spectrum of the massless state
as in the case of the cross spectrum.

\subsection{Power-spectra in the flavor eigenstates}

We calculate the tensor spectra in the flavor eigenstates by making use
of the results in the previous subsection. We will see all of them
agree with each other and these amplitudes are conserved on
super-horizon scales. We also see
the spectral index and the amplitudes in the leading order.

Using the relation between the mass eigenstates and the flavor eigenstates:
\begin{eqnarray}
\begin{pmatrix}
 q \\ p
\end{pmatrix}
=
\frac{1}{(\kappa^2+\epsilon^2)^{1/2}}
\begin{pmatrix}
 \kappa & -\epsilon \\ \kappa & \kappa^2/\epsilon
\end{pmatrix}
\begin{pmatrix}
 x \\ y
\end{pmatrix}\ ,
\end{eqnarray}
we can obtain the following power-spectra
\begin{eqnarray}
 \langle qq \rangle 
= \frac{1}{\kappa^2+\epsilon^2}\bigl[\kappa^2\langle xx \rangle -\kappa \epsilon(\langle xy \rangle+\langle yx \rangle) +\epsilon^2 \langle yy \rangle\bigr] \ ,
\end{eqnarray}
\begin{eqnarray}
 \langle qp \rangle 
= \frac{1}{\kappa^2+\epsilon^2}\bigl[\kappa^2\langle xx \rangle +(\kappa^3/ \epsilon)\langle xy \rangle-\kappa\epsilon\langle yx \rangle -\kappa^2 \langle yy \rangle\bigr]\ ,
\end{eqnarray}
\begin{eqnarray}
 \langle pp \rangle 
= \frac{1}{\kappa^2+\epsilon^2}\bigl[\kappa^2\langle xx \rangle +(\kappa^3/ \epsilon)(\langle xy \rangle+\langle yx \rangle) + (\kappa^4/\epsilon^2) \langle yy \rangle\bigr]\ .
\end{eqnarray}
According to the results in the previous subsection, we find
\begin{eqnarray}
 \langle qq \rangle &=& \langle qp \rangle = \langle pp \rangle \nonumber \\
 &=& \frac{\kappa^2}{\kappa^2+\epsilon^2}\langle xx \rangle +{\mathcal O}(\eta^3) \nonumber \\
&=& \int d(\log k) \frac{\kappa^2}{\kappa^2+\epsilon^2}\biggl(\frac{H_0}{\pi
  M_g}\biggr)^2
  (-H_0\eta)^{-2s\frac{2\epsilon_0^2(1-\epsilon_0)}{\kappa_0^2+\epsilon_0^2}}\biggl(\frac{k}{2H_0}\biggr)^{-2s\bigl(1+\frac{2\epsilon_0^2(1-\epsilon_0)}{\kappa_0^2+\epsilon_0^2}\bigr)}\nonumber  \\
&&\qquad \qquad \qquad \qquad \qquad \qquad \qquad \qquad \times \biggl(\frac{\Gamma(\nu_X)}{\Gamma(\frac{3}{2})}\biggr)^2\biggl(1-\frac{6s\epsilon_0^2(1-\epsilon_0)}{\kappa_0^2+\epsilon_0^2}\biggr)\ .
\label{index}
\end{eqnarray}
The power-spectrum of the physical metric, that of the other metric and the
cross spectrum are the same in the first order of the slow-roll
parameter at the end of inflation.
When we include both plus mode and cross mode, the results should be multiplied by two.
Superficially, these spectra do not seem to conserve.
However, we can verify their conservation as follows
\begin{eqnarray}
 \frac{{\rm d}\log \langle q^2 \rangle}{{\rm d}\eta}
&=&\frac{{\rm d}}{{\rm d}\eta}\biggl[-\log (\kappa^2+\epsilon^2)
+\Bigl(-2s\frac{2\epsilon_0^2(1-\epsilon_0)}{\kappa_0^2+\epsilon_0^2}\Bigr)\log(-\eta)\biggr]
\nonumber \\
&=&-\frac{2\epsilon\epsilon^{\prime}}{\kappa^2+\epsilon^2}-2s\frac{2\epsilon_0^2(1-\epsilon_0)}{\kappa_0^2+\epsilon_0^2}\frac{1}{\eta} \nonumber \\
&=&-\frac{-4s{\mathcal H}\epsilon_0^2(1-\epsilon_0)}{\kappa_0^2+\epsilon_0^2}-\frac{4s\epsilon_0^2(1-\epsilon_0)}{\kappa_0^2+\epsilon_0^2}\frac{1}{\eta}\nonumber \\
&=&0 \ .
\end{eqnarray}
We can read off the spectral index of the tensor power-spectrum from eq.(\ref{index}) as
\begin{eqnarray}
 n_T=-2s\Bigl(1+\frac{2\epsilon_0^2(1-\epsilon_0)}{\kappa_0^2+\epsilon_0^2}\Bigr) \ .
\end{eqnarray}
It means that the spectra are red-tilted since $\epsilon_0$ has a value
between $0$ and $1$ as we mentioned in section \ref{sec.IBS}.
The amplitudes of them in the leading order, i.e. in the slow-roll limit, are
\begin{eqnarray}
 \langle qq \rangle &=& \langle qp \rangle = \langle pp \rangle 
= \frac{\kappa_0^2}{\kappa_0^2+\epsilon_0^2}\biggl(\frac{H_0}{\pi
  M_g}\biggr)^2 \ .
\end{eqnarray}
The amplitude of the physical metric is suppressed compared with that in the conventional case 
since the physical metric is represented as 
a superposition of the massless mode and the massive mode and the massive mode rapidly decays.
When we take the general relativity limit ($M_f/M_g\rightarrow 0$),
the amplitude and the spectral index of the physical metric are smoothly
connected to those in the conventional case since $\kappa_0$ becomes infinity.
On the other hand, when we take the massive gravity limit ($M_f/M_g\rightarrow \infty$),
the amplitude vanishes since $\kappa_0$ becomes zero.

We emphasize that the physical tensor modes and the other tensor modes are
maximally correlated and they identically behave at the end of inflation.

\section{Conclusion}\label{sec.Con}

The deviation of the tensor power-spectrum from the conventional case in the slow-roll limit 
is that the total amplitude is suppressed due to
the mixing of the physical metric and the other metric.
The spectral index is red-tilted compared with the conventional case in the leading order.
The red-tilted spectrum originates in the time dependence of the
mixing matrix, i.e. graviton oscillation. 
The difference between the lapse functions results in that the
propagation speed of the massless state is larger than the speed of light
and the amplitude is modified in the first order of the slow-roll parameter.
We note that the physical metric and the other metric are maximally
correlated.
This indicates that we can find some information about the other metric by
observing physical quantities.

It would be easy to extend the analysis to multi-metric gravity \cite{Khosravi:2011zi,Nomura:2012xr}.
We have to calculate not only the tensor perturbations but also the scalar perturbations
in order to connect these results to observational data such as the tensor-to-scalar ratio.
Though someone worries that this branch may suffer gradient instability during the radiation dominant era
\cite{Comelli:2012db, DeFelice:2014nja},
gradient instability will be avoided when we consider the bare mass is not so small 
to explain the current accelerated expansion. 
This situation is allowed since we have not only a massive graviton but also a massless graviton
in the case of bimetric gravity.
Other possibility is that we can obtain another branch which has no Higuchi ghost
when we extend the minimal model to general models including other parameters.
In those cases, we may need to have some extreme values for the theoretical parameters to obtain the new branch.
We are also interested in the behavior of the tensor perturbations 
in the reheating era. It is because the other metric
interacts with the scalar field in a non-trivial way through $\zeta$.
The function $\zeta$ oscillates when the scaler field oscillates
and this will cause the parametric amplification of the other metric.
Then, the physical metric can be enhanced through the mixing of the
physical metric and the other metric.

%===============================================================%
%******************** Acknowledgements *************************%
%===============================================================%
\section*{Acknowledgement}
We would like to thank Antonio de Felice, Claudia de Rham, Tetsuya
Shiromizu and Takahiro Tanaka for fruitful discussions. 
YS is supported by the Grant-in-Aid for Japan Society for the Promotion
of Science(JSPS) Fellows No. 261236.
JS is supported by Grants-in-Aid for Scientific Research (C) No.25400251 and Grants-in-Aid for 
Scientific Research on Innovative Areas No.26104708.

%

%\bibliographystyle{apsrev4-1}
%\bibliography{references}

\end{document}